# Validation of gait characteristics extracted from raw accelerometry during walking against measures of physical function, mobility, fatigability, and fitness

Jacek K. Urbanek[1], Vadim Zipunnikov[1], Tamara Harris[2],

Ciprian Crainiceanu[1], Jaroslaw Harezlak[3], Nancy W. Glynn[4]

1) Department of Biostatistics, Bloomberg School of Public Health, Johns Hopkins University, 615 N. Wolfe Street, Baltimore, MD

2) Laboratory of Epidemiology, Demography, and Biometry, National Institute on Aging

3) Department of Epidemiology and Biostatistics, School of Public Health, Indiana University, Bloomington, IN

4) Center for Aging and Population Health, Department of Epidemiology, Graduate School of Public Health, University of Pittsburgh, Pittsburgh PA

Address correspondence to Jacek K. Urbanek:

Department of Biostatistics, Bloomberg School of Public Health, Johns Hopkins University, 615 N. Wolfe Street, Baltimore, MD, email: jurbane2@jhu.edu

**ABSTRACT**

***Background.*** Data collected by wearable accelerometry devices can be used to identify periods of sustained harmonic walking. This report aims to establish whether the features of walking identified in the laboratory and free-living environments are associated with each other as well as measures of physical function, mobility, fatigability, and fitness.

***Methods.*** Fifty-one older adults (mean age 78.31) enrolled in the Developmental Epidemiologic Cohort Study were included in the analyses. The study included an "in-the-lab" component as well as 7 days of monitoring "in-the-wild" (free-living). Participants were equipped with hip-worn Actigraph GT3X+ activity monitors, which collect raw accelerometry data. We applied a walking identification algorithm and defined features of walking, including participant-specific walking acceleration and cadence. The association between these walking features and physical function, mobility, fatigability, and fitness was quantified using linear regression analysis.

***Results.*** Acceleration and cadence estimated from "in-the-lab" and "in-the-wild" data were significantly associated with each other ($p < 0.05$). However, walking acceleration "in-the-lab" were on average 96% higher than "in-the-wild", whereas cadence "in-the-lab" was on average 20% higher than "in-the-wild"**.** Acceleration and cadence were associated with measures of physical function, mobility, fatigability, and fitness ($p < 0.05$) in both "in-the-lab" and "in-the-wild" settings. Additionally, "in-the-wild" daily walking time was associated with fitness ($p < 0.05$).

***Conclusions.*** The quantitative difference in proposed walking features indicates that participants may over-perform when observed "in-the-lab". Also, proposed features of walking were significantly associated with measures of physical function, mobility, fatigability, and fitness, which provides evidence of convergent validity.

**INTRODUCTION**

The use of accelerometry has become popular in aging studies (1, 2, 3). Features derived from data collected by wearable accelerometers "in-the-lab" (or clinic) environment are often used as practical measures of mobility (4) or fatigability (5). Advanced data processing algorithms allow the analysis of specific characteristics of human gait, such as cadence and asymmetry (6). While "in-the-lab" experiments have significant potential for medical and epidemiological studies, there is an increased interest in the data collected "in-the-wild", that is, in the free-living environment. Several large–scale studies, including the National Health and Nutrition Examination Survey (NHANES) (7), the Baltimore Longitudinal Study on Aging (8) and the Women's Health Initiative (WHI) (9), have either collected or are in the process of collecting "in-the-wild" activity data. As levels of activity observed "in-the-wild" have been shown to be highly correlated with both health and aging outcomes (10, 11), there is an increased interest in evaluating additional measures of activity, especially using the raw, high resolution data.

Early research on physical activity focused on aggregated measures of activity in one-minute epochs. This was due to both hardware and software limitations. Thus, most published evidence is based on a particular type of aggregation of information at the minute level. This raises questions about whether accumulating data differently can extract additional layers of information. Current accelerometry data are routinely recorded at the sub-second level, usually between 10 and 100 observations per second. For example, the most recent NHANES cohort (2013 – 2014) and the WHI study collect accelerometry data at the sub-second level. Such data offer the promise for more detailed information that can be extracted using specialized analytic approaches. Some promising early studies indicate that types of activity can be predicted at the sub-minute level (12). In this paper, we focus on estimating gait characteristics based on "in-the-lab" and "in-the-wild" sub-second accelerometry data. Gait has been shown to be associated with longevity (13), obesity (14) and progression of Parkinson disease (15).

We use a previously developed algorithm for detection of sustained harmonic walking (SHW) (16) to extract gait characteristics in every period identified as SHW. Our aims included examining comparable gait characteristics collected in the controlled ("in-the-lab") and uncontrolled ("in-the-wild") environment and to examine the associations between accelerometry-derived gait characteristics with common measures of physical function, mobility, fatigability, and fitness to establish convergent validity of our novel gait characteristics. Additionally, we investigate the potential of high-density data collected in controlled ("in-the-lab") and uncontrolled ("in-the-wild") environment in epidemiological and clinical research. To achieve these goals we compare the predictive power of gait characteristics derived from both conditions. We hypothesize that: 1) features of SHW obtained from high-density accelerometry "in-the-lab" differ from features of the SHW obtained from high-density accelerometry "in-the-wild" 2) features of SHW obtained from high-density accelerometry "in-the-lab" data will be associated with measures of physical function, mobility, fatigability, and fitness; and 3) features of the SHW obtained from high-density accelerometry "in-the-wild" data will also be associated with the same physical performance measures.

**METHODS**

***Study participants***

Eighty-nine community-dwelling older adults were recruited from the Pittsburgh, Pennsylvania area for the National Institute on Aging, Aging Research Evaluating

Accelerometry (AREA) project, part of the Developmental Epidemiologic Cohort Study (DECOS) (17). AREA was a cross-sectional methodological initiative designed to examine the impact of accelerometry wear location on assessment of physical activity and sedentary behavior among 89 older adults enrolled between March and May of 2010. This report included data from 51 healthy participants (25 men and 26 women) who were available to us and had complete "in-the-lab" (N=46) or "in-the-wild" (N=48) accelerometry data. Individuals were excluded from the DECOS study if they suffered from any of the following conditions: hip fracture, stroke in the past 12 months, cerebral hemorrhage in the past 6 months, heart attack, angioplasty, heart surgery in the past 3 months, chest pain during walking in the past 30 days, current treatment for shortness of breath or a lung condition, usual aching, stiffness, or pain in their lower limbs and joints and bilateral difficulty bending or straightening the knees fully (18).

*Accelerometry Measures*

All participants were equipped with Actigraph GT3X+ accelerometers placed on the right hip during both the "in-the-lab" and "in-the-wild" experiments. Devices collected raw accelerometry data along three orthogonal axes with the sampling frequency of 80 observations-per-second (80Hz). For the "in-the-lab" condition, participants wore the accelerometer during a 4-meter walk and two 400-m walks. During the "in-the-wild" condition, participants were equipped with the accelerometer for seven consecutive days and were told to maintain their normal, unsupervised, free-living activities. They were instructed to take off the activity monitor only during sleep. Compliance was addressed via visual examination of the data. All participants included in the experiment adhered to the protocol.

Periods of SHW were labeled using a walking algorithm detection based on the tri-axial accelerometry data ( 16). The SHW detection algorithm returns the temporal location and duration of walking bouts based on the raw accelerometry data. Additionally, the vector magnitude count (VMC) and cadence (expressed in steps-per-minute) are estimated for each walking bout. VMC is defined as the mean absolute deviation of the acceleration signal produced during SHW and is expressed in standard gravity (g = 9.81 m/s2) units ( 16). For each estimated walking bout we computed a set of walking features including duration, VMC and cadence, for both "in-the-lab" and "in-the-wild" data. To ensure the stability of these measures, we used only SHW bouts longer than or equal to 20 seconds (30% of all identified bouts of walking). We defined micro-scale features of walking as characteristics describing mechanics of human gait. In this report we use: walking VMC (expressed in g) and cadence (steps/second). Analogously, we define macro-scale features of walking as characteristics representing overall volume of walking "in-the-wild". Here, as a macro-scale feature, we use average daily walking time (expressed in minutes per day).

For data collected during the controlled laboratory experiment we derived the median acceleration and the median cadence during SHW. The median acceleration was defined as the median of all VMCs during SHW of one individual. The median cadence, expressed in steps per minute, was defined as the median of estimated cadence during SHW bouts. For data collected "in-the-wild", we derive the same micro-scale gait dynamics: median acceleration and median cadence. Additionally, we quantified the average daily walking time as the total estimated time of SHW (expressed in minutes) divided by the number of days. The average daily acceleration was obtained by dividing the total VMC during the monitoring period by the number of monitoring days. These variables depend on the total estimated SHW time and reflect the overall macro-scale activity of the individuals.

*Outcomes of interest*

We focused on several outcomes of interest representing different aspects of physical performance of older adults. Physical function was assessed using the Short Physical Performance Battery (SPPB) and included standing balance, chair stands, and a 6 meter usual-paced walk ( 19). Each component had a possible score of 0-4. Total SPPB scores ranged from 0-12, with higher scores indicating better physical function. Mobility was assessed by the fastest of two, usual-paced 6m walking tests from the SPPB ( 19) as well as time to finish a usual-paced 400m walk ( 20). Fitness was considered as time to complete a fast-paced 400m walk ( 21).

For both 400m walks, the course was set-up in a long, secured hallway with markers at both ends spaced by 20 meters. For safety purposes participants wore a heart rate monitor (Polar Chest Transmitter, Warminster, PA). The 400m walk test had two parts: a warm-up at usual pace (2 laps) and then either the 400m usual or fast walk test (10 laps). Participants were excluded if, after the warm-up, their resting heart rate was over 110 or under 40 beats per minute, or if they had systolic blood pressure higher than 200 mmHg or diastolic blood pressure higher than 110 mmHg. For the fast-paced 400m walk, participants were told to walk as quickly as possible without running at a pace they could maintain for ten laps. The usual-paced 400-m walk was administered during the second clinic visit in an identical manner to the fast-paced walk with the exception that participants were instructed to walk at their usual, normal pace during the 400m walk portion ( 17). All participants were able to complete the entire sequence of activities, which included several non-walking activities as well as resting periods. Trained research personnel administered the tests and labeled periods that corresponded to walking.

Perceived physical fatigability was measured using the Pittsburgh Fatigability Scale (PFS), a 10-item self-administered questionnaire anchoring fatigue to intensity and duration of common activities performed by older adults ( 22).  Summary scores on the PFS range from 0-50 with higher scores denoting higher fatigability.

### Other measures

Age, sex, race, and education were ascertained by questionnaire.  Body mass index (BMI) was calculated in weight in kilograms per squared height in meters using a stadiometer and a standard physician's balance scale. Self-report physical activity was measured using the Community Healthy Activities Model Program for Seniors (CHAMPS) questionnaire ( 23).  All measures were collected during Clinic Visit 1.

### Statistical methods

To investigate the relation between features of walking, outcomes of interest and demographic characteristics, we include correlation matrix of the bivariate associations among the study variables (Table 3).  The primary analysis consisted of determining predictive factors for each of the measures of physical function (SPPB), mobility (gait speed and usual-paced 400 meter walk time), physical fatigability and fitness (fast-paced 400 meter walk time). We fit the linear regression models for each of the aforementioned outcomes separately.  We consider models where the candidate predictors were activity-derived factors, walking-derived factors and demographic information including age, sex, height and BMI. We use separate models for:  1) "in-the-lab" walking-derived factors: median walking acceleration and median cadence; and 2) "in-the-wild" activity-derived factors: daily acceleration, daily walking time, median walking acceleration, and median cadence.  Within each linear regression model (5 outcomes and 2 predictor types: "in-the-lab" and "in-the-wild") the candidate markers are evaluated using a stepwise model selection procedure based on the Akaike's Information Criterion (AIC) (24). Selected sparse models represent the optimal combination of predictors for given outcome of interest.  Final models' prediction performance is assessed by the adjusted R-squared.

**RESULTS**

Demographic characteristics of the study participants are summarized in Table 1. All summary statistics are reported using means, medians and interquartile ranges.  Mean age was 78 years and the mean BMI was 26.8 kg/m$^2$.  The cohort was generally high functioning with the mean SPPB score 10.4 (Q1=10.0, Q2=11.0, Q3=12.0) and mean usual gait speed 1.15 meters/second (Q1=1.03, Q2=1.11, Q3=1.23). The mean time to complete the usual-pace 400 meter walk was 383.1 seconds (Q1=352.7, Q2=379.4, Q3=414.8) while for fast-pace 400 meter walk it was 329.4 seconds (Q1=278.9, Q2=329.0, Q3=355.5).   The mean Pittsburgh Fatigability Scale physical fatigability score was 16.1 (Q1=11.5, Q2=14.0, Q3=20.5), denoting higher fatigability.   Self-report moderate intensity physical activity for this group exceeded the AHA/ACSM recommended guidelines of 150 min/week with a median of 310 min/wk, but with a wide range of values.

Descriptive characteristics for the micro- and macro-scale walking features from the "in-the-lab" and "in-the-wild" can be found in Table 2. Average median walking acceleration "in-the-lab" was 0.2442 g (Q1=0.1831, Q2=0.2254, Q3=0.3014) while average median cadence "in-the-lab" was 122.76 (Q1=117.0, Q2=124.0, Q3=129.5) steps per minute.  Average median walking acceleration "in-the-wild" was 0.12457 (Q1=0.09590, Q2=0.12982, Q3=0.14784) while average median cadence "in-the-wild" was 101.460 (Q1=93.56, Q2=101.50, Q3=106.25) steps per minute. Walking acceleration "in-the-lab" were on average 96% higher than "in-the-wild", whereas cadence "in-the-lab" was on average 20% higher than "in-the-wild".

Correlation between micro-scale features of walking observed "in-the-wild" and "in-the-lab" was relatively high for walking acceleration (r = 0.72) and lower for cadence (r = 0.45). Correlation between walking acceleration and cadence was 0.58 for "in-the-lab" and 0.7 for "in-the-wild" setting. Micro-scale features of walking were significantly correlated with all outcomes of interest for both "in-the-lab" and "in-the-wild" settings except correlation between median cadence "in-the-wild" and mobility (time to finish usual-paced 400 meter walk). Correlations between features of walking and outcomes of interest were generally higher for "in-the-lab" settings except for fatigability where correlation was higher for walking observed "in-the-wild". Macro-scale feature of walking (daily walking time) was significantly correlated only with fitness (time to finish fast-paced 400 meter walk). Daily acceleration was significantly correlated with all outcomes of interest except fatigability. All bivariate associations among the study variables are included in Table 3.

Estimated normalized regression coefficients and their corresponding p-values (in brackets) for "in-the-lab" data are shown in Table 4. For the "in-the-lab" models, median acceleration produced during fast-paced 400-meter walk had a strong  (p < 0.005) statistical association with physical function ($\beta$ = 0.497, p < 0.001) and fatigability ($\beta$ = -0.422, p = 0.003). Median walking acceleration was also strongly associated with mobility ($\beta$ = 0.381, p = 0.003 for gait speed and $\beta$ = -0.636, p < 0.001 for time to finish usual-paced 400m walk) and fitness ($\beta$ = -0.607, p < 0.001).  Median cadence was significantly (p < 0.05) associated with mobility ($\beta$ = 0.298, p = 0.022 for gait speed and $\beta$ = -0.234, p = 0.034 for time to finish usual-paced 400m walk) and fitness ($\beta$ = -0.215, p = 0.049 for time to finish fast-paced 400m walk time).

For the "in-the-wild" models (Table 5), median walking acceleration was strongly statistically associated with physical function ($\beta$ = 0.332, p = 0.001) and fatigability ($\beta$ = -0.551, p < 0.001). Median cadence was statistically associated with mobility ($\beta$ = 0.378, p = 0.002 for gait speed and $\beta$ = -0.799, p = 0.034 for time to finish usual-paced 400m walk time) and fitness ($\beta$ = -0.378, p = 0.002 for time to finish fast-paced 400m walk time). Daily walking time was strongly associated with fitness ($\beta$ = -0.318, p = 0.004). This association did not extend to daily acceleration.

**DISCUSSION**

Using recently developed methodology for identification of bouts of walking (16), we derived measures representing both macro-scale overall volume of daily walking in a free-living environment (daily walking time) and micro-scale gait properties (median walking acceleration and median cadence) for both "in-the-lab" and "in-the-wild" conditions. We observed that micro-scale features of walking "in-the-lab" are on average higher than "in-the-wild". Additionally, we have found that among healthy older adults accelerometry-derived features of SHW are significantly associated with measures of physical function, mobility, fatigability, and fitness, demonstrating the potential of the raw accelerometry data to be used as a novel source of information characterizing physical function, mobility, fatigability, and fitness.

Results of analysis of bivariate associations between study variables were consistent with the results from the linear regression models. Interestingly, we observed relatively low correlations between cadence "in-the-lab" and "in-the-wild". Also, cadence "in-the-lab" was on average higher than cadence "in-the-wild". Similar patterns could be observed for walking acceleration where, despite a high correlation, values observed "in-the-lab" where on average higher than their "in-the-wild" counterparts. We identify two main reasons for this difference. First, because the "in-the-lab" condition was conducted in the morning participants were relatively unfatigued, therefore were able to perform on higher level. Second, participants were more likely to over-perform, as they were fully aware of the purpose of this study.

Results of regression analyses presented in this paper complement previously published findings for controlled "in-the-lab" experiments. For example, a significant positive association between amplitude of walking acceleration and gait speed has been previously reported for "in-the-lab" experiments (25, 26). However, this work is novel as it extended those findings by demonstrating an association between median cadence and gait speed "in-the-wild".

Micro-scale features of walking (acceleration and cadence) were associated with physical function, mobility, fatigability, and fitness. In addition, we found that "in-the-lab" median walking acceleration was associated with gait speed. Thus, median walking acceleration complements median cadence as a predictor of gait speed.

In general, micro-scale features of walking "in-the-lab" were more closely related to outcomes of interest than features of walking "in-the-wild" (adjusted $R^2$ values in Table 4 and Table 5). The only exception was physical fatigability, where median walking acceleration "in-the-lab" was able to explain more variability than walking acceleration "in-the-wild". This finding suggests that physical fatigability can be better identified in an free-living, unconstrained gait parameters as oppose to "in-the-lab" conditions where participants were well rested which may have positively influenced their physical performance.

The work presented in this paper consisted of a small number of healthy participants with relatively high levels of self-reported physical activity. To fully understand the generalizability of these findings, proposed methodology should be extended to a larger sample representing a wider range of function, including people with sedentary lifestyle, gait impairments and the need for walking-aids.

Despite these limitations we believe that this work is a step towards wearable accelerometers being used in the future as a reliable method for monitoring physical function, mobility, fatigability, and fitness "in-the-wild", allowing for less-costly and patient friendly measurements. Objective data collection using modern, small-size wearable devices is indeed easy to implement, inexpensive and non-invasive. With increasing access to new technologies, wearable activity monitors could become standard in modern health research and, more importantly, in modern healthcare, similar to Holter monitors and DEXCOM continuous glucose monitors ( 27, 28). Continuous observation of patients' physical activity levels together with micro- and macro-scale features of walking could, for example, signal early detection of decline in physical performance and fatigability so it can be addressed before the downward spiral of disability. Additionally, wearable technology will make it possible to better identify patients' response to a therapy or interventions aimed at increasing physical activity.

In conclusion, our findings indicate that both "in-the-lab" and "in-the-wild" conditions provide independent information about the physical performance of older adults. Data collected "in-the-wild" reflect more natural, usual state of participants; while data collected "in-the-lab" are indicative of higher-level, unfatigued performance. Analysis of within-subject differences in performance "in-the-lab" and "in-the-wild" and their relation to health and physical function will be examined in future work. Data collected "in-the-lab" are generally the main source of information in modern epidemiological studies on physical performance. Our results indicate that micro- and macro-scale gait parameters can be extracted and quantified from data collected in modern accelerometry studies both "in-the-lab" and "in-the-wild". The amount of data collected

"in-the-wild" is typically much larger, as it can be continuously measured for weeks or months at a time. Here, we propose to interpret features of SHW observed "in-the-wild" as an objective and unbiased proxies of physical function, mobility, fatigability, and fitness. Indeed, even the best-designed "in-the-lab" experiment cannot fully capture the natural free-living environment conditions. By collecting data "in-the-wild" we avoid potential biases introduced by one's tendency to under- or over-perform during supervised in-lab experiments. Although participants were aware of the monitoring device, which could have influenced their free-living behavior, we believe that the 7 day monitoring period eliminates any potential biases as well as influence of many uncontrollable factors, including type of walking surface, elevation, and environmental conditions ( 29). Future research will focus on developing methodology that could serve as an alternative for "in-the-lab" measures of physical function, mobility, fatigability, and fitness using features of physical activity estimated only with "in-the-wild" data as well as extending this work to validate wrist-worn devices.

## Acknowledgements

This research was supported by Pittsburgh Claude D. Pepper Older Americans Independence Center, Research Registry, and Developmental Pilot Grant (PI: Glynn) – NIH P30 AG024826 and NIH P30 AG024827. National Institute on Aging Professional Services Contract HHSN271201100605P. NIA Aging Training Grant (PI: AB Newman) T32-AG-000181. The project was supported, in part, by the Intramural Research Program of the National Institute on Aging. This research was supported by the NIH grant RO1 NS085211 from the National Institute of Neurological Disorders and Stroke, by the NIH grant RO1 MH095836 from the National Institute of Mental Health and by NIH grant R01 HL123407 from National Heath, Lung and Blood Institute. Dr Harezlak's research was supported in part by the NIMH grant R01 MH108467.

*Table 1. Characteristics of study population (N = 51)*

| Variable | Mean (Q1, Q2, Q3) or N (%) |
|---|---|
| Age [yr.] | 78.31 (74.0, 77.5, 82.0) |
| Sex (Male) | 25 (49%) |
| BMI | 26.79 (23.6, 25.9, 30.0) |
| Height [cm] | 165.7 (159.5, 166.3, 171.9) |
| Race: | |
|     White | 46 (90%) |
|     Black | 4 (8%) |
|     Asian | 1 (2%) |
| Education: | |
|     High School | 9 (18%) |
|     College | 25 (49%) |
|     Graduate | 17 (33%) |
| Self-report moderate intensity activity [min./week] | 310.1 (75.0, 232.5, 397.5) |
| SPPB score | 10.4 (10.0, 11.0, 12.0) |
| Usual gait speed [meters/second] | 1.15 (1.03, 1.11, 1.23) |
| Usual 400m walk time [second] | 383.1 (352.7, 379.4, 414.8) |
| Physical fatigability score | 16.1 (11.5, 14.0, 20.5) |
| Fast 400m walk time [second] | 329.4 (278.9, 329.0, 355.5) |

*Table 2. Means and values of Q1, Q2 and Q3 (in the brackets) for the micro- and macro-scale walking features.*

| Variable | Mean (Q1, Q2, Q3) or N (%) | |
| --- | --- | --- |
| | "in-the-lab" | "in-the-wild" |
| Median Walking Acceleration [g] | 0.2442 (0.1831, 0.2254, 0.3014) | 0.12457 (0.09590, 0.12982, 0.14784) |
| Median Cadence [steps/min.] | 122.76 (117.0, 124.0, 129.5) | 101.460 (93.56, 101.50, 106.25) |
| Daily Walking Time [min.] | | 59.502 (38.712, 60.334, 81.244) |
| Daily Acceleration [g] | | 77970 (51387, 71273, 91565) |

*Table 3. Pearson correlation matrix of the bivariate associations among the study variables.*

| | 1 | 2 | 3 | 4 | 5 | 6 | 7 | 8 | 9 | 10 |
|---|---|---|---|---|---|---|---|---|---|---|
| 1) SPPB | | | | | | | | | | |
| 2) 6m gait speed | .41* | | | | | | | | | |
| 3) Usual-paced 400m walk | -.49* | -.64* | | | | | | | | |
| 4) Physical Fatigability | -.38* | -.40* | .47* | | | | | | | |
| 5) Fast-paced 400m walk | -.47* | -.49* | .85* | .37* | | | | | | |
| 6) Median walking acc. "in-the-lab" | .53* | .55* | -.75* | -.42* | -.78* | | | | | |
| 7) Median cadence "in-the-lab" | .41* | .44* | -.66 | -.34* | -.61* | .58* | | | | |
| 8) Daily acc. | .37* | .34* | -.53* | -.20 | -.628 | .77* | .28 | | | |
| 9) Daily walking time | .15 | -.07 | -.22 | .22 | -.35* | .26 | .31* | .50* | | |
| 10) Median walking acc. "in-the-wild" | .46* | .44* | -.45* | -.55* | -.58* | .72* | .29 | .60* | -.09 | |
| 11) Median cadence "in-the-wild" | .31* | .39* | -.26 | -.46* | -.34* | .40* | .45* | .35* | -.01 | .70* |

*p < 0.05

*Table. 4. Estimated normalized regression coefficients and the corresponding p-values (in brackets) of the best-fitted models for data collected "in-the-lab".*

| | Physical Function (SPPB) | Mobility (6m Gait speed) | Mobility (Usual-paced 400m walk) | Physical Fatigability | Fitness (Fast-paced 400m walk) |
|---|---|---|---|---|---|
| Median walking acceleration | 0.497 (<0.001) | 0.381 (0.003) | - 0.636 (<0.001) | - 0.422 (0.003) | - 0.607 (<0.001) |
| Median cadence | | 0.298 (0.022) | - 0.234 (0.034) | | - 0.215 (0.049) |
| Age | | | | | |
| Height | | 0.479 (<0.001) | - 0.511 (<0.001) | | |
| BMI | | | | | |
| Sex (Female) | | | - 0.733 (0.011) | | |
| Adjusted $R^2$ | 0.26 | 0.52 | 0.71 | 0.18 | 0.63 |

*Table. 5. Estimated normalized regression coefficients and the corresponding p-values (in brackets) of the best-fitted models for data collected "in-the-wild".*

| | Physical Function (SPPB) | Mobility (6m gait speed) | Mobility (Usual-paced 400m walk) | Physical Fatigability | Fitness (Fast-paced 400m walk) |
|---|---|---|---|---|---|
| Daily acceleration | | | | | |
| Daily walking time | | | - 0.255 (0.048) | | - 0.318 (0.004) |
| Median walking acceleration | 0.332 (0.001) | | - 0.799 (<0.001) | -0.551 (<0.001) | -0.378 (0.002) |
| Median cadence | | 0.378 (0.002) | - 0.362 (0.034) | | |
| Age | | | | | 0.344 (0.013) |
| Height | | 0.3324 (0.005) | - 0.548 (0.004) | | |
| BMI | | | | | 0.263 (0.015) |
| Sex (Female) | | | - 1.267 (0.003) | | |
| Adjusted $R^2$ | 0.19 | 0.27 | 0.38 | 0.30 | 0.56 |